\def\aa{{AA}}
\def\aj{{AJ}}

\def\apj{{ApJ}}
\def\apjl{{ApJ}}
\def\apjs{{ApJS}}

\def\mnras{{MNRAS}}

\def\sol{\ifmmode _{\mathord\odot}
         \else $_{\mathord\odot}$\fi}

\documentstyle[12pt,aasms4]{article}

\begin{document}
\baselineskip 0.6 true cm

\lefthead{A. S. Cohen, et al.}

\righthead{Delay, Magnification, and Variability in 0218+357}

\title{FURTHER INVESTIGATION OF THE TIME DELAY, MAGNIFICATION RATIOS,
AND VARIABILITY IN THE GRAVITATIONAL LENS 0218+357}
\author{A. S. Cohen, J. N. Hewitt} 
\affil{Department of Physics and Center for Space Research \\
Room 37-607, Massachusetts Institute of Technology, Cambridge, MA  02139 \\
aaron@space.mit.edu, jhewitt@mit.edu}
\author{C. B. Moore}
\affil{Harvard-Smithsonian Center for Astrophysics, 60 Garden Street \\
 Cambridge, MA 02139 \\ cmoore@cfa.harvard.edu}
\author{D. B. Haarsma}
\affil{Department of Physics and Astronomy, Calvin College 
\\ Grand Rapids, MI  49546 \\ dhaarsma@calvin.edu}
\and

\begin{abstract}
High precision VLA flux density measurements for the lensed images
of 0218+357 yield a time delay of $10.1^{+1.5}_{-1.6}$ days
(95\% confidence).
This is consistent with independent measurements carried out at
the same epoch (\cite{biggs99}), lending confidence in the robustness
of the time delay measurement.
However, since both measurements make use of the same features
in the light curves, it is possible that the effects
of unmodelled processes, such
as scintillation or microlensing, are biasing both time delay
measurements in the same way.
Our time delay estimates result in a confidence intervals
that are somewhat larger than those of Biggs et al., probably
because we adopt a more general model of the source variability, allowing
for constant and variable components.
When considered in relation to the lens mass model of Biggs et al.,
our best-fit time delay implies a Hubble constant of
$H_o = 71^{+17}_{-23}$~km~s$^{-1}$~Mpc$^{-1}$
for $\Omega_o = 1$ and $\lambda_o=0$
(95\% confidence; filled beam).
This confidence interval for $H_o$ does not reflect systematic error,
which may be substantial, due to uncertainty in the position of the
lens galaxy.
We also measure the flux ratio of the {\it variable} components
of 0218+357, a measurement of a small region that should more closely
represent the true lens magnification ratio.
We find ratios of $3.2^{+0.3}_{-0.4}$ (95\% confidence; 8~GHz) and
$4.3^{+0.5}_{-0.8}$ (15~GHz).
Unlike the reported flux ratios on scales of $0.1''$, these ratios
are not strongly significantly different.
We investigate the significance of apparent differences
in the variability  properties of the two images of the background
active galactic nucleus.  We conclude that
the differences are not significant, and that time series much longer
than our 100-day time series will be required to investigate
propagation effects in this way.
\end{abstract}


\section{Introduction}

Gravitational lenses provide unique opportunities to make 
cosmological measurements and to study distant sources and lensing
galaxies.
One such measurement is that
of cosmological distance, without the assumptions
inherent in intermediate distance indicators (\cite{refsdal64}; \cite{refsdal66}).
The difference in the light travel time of the
paths to the images is proportional to the angular diameter distance
to the lens (if the mass scale in the lens can be determined independently)
or to an ``effective'' distance, a combination of
the angular diameter distances between the lens, observer, and source
(\cite{narayan91}).
The difference in the light travel time can be found
from the time
delay between flux variations in each lensed image.
A mass model predicts the relationship between time delay and angular
diameter distance.
Thus, a time delay and a mass model can be used to determine independently
the angular diameter distance to the lens.
Another prediction of a gravitational lens model is the magnification
ratio.
This quantity can be used as a model constraint or as a means
to investigate source structure on very small scales (see, for example,
\cite{conner92}).
Finally, studies of the variability of lensed images,
particularly any differences in the variability properties, present a new
tool for investigating propagation effects.

Gravitational lens
monitoring observations were first carried out for 
the gravitational lens 0957+561, with
time delays reported by Florentin-Nielsen (1984), Vanderriest et al.
(1989), Leh\'ar et al. (1992), Press, Rybicki, and Hewitt (1992a\&b)
and Oscoz et al. 1997, among others.
(For a comprehensive review of time delay measurements in 0957+561 see
\cite{haarsma97}.)
The results were conflicting and somewhat controversial, and the
discrepancies have only 
recently been resolved (\cite{kundic97}, \cite{haarsma99}).
Though these measurements, particularly those of Kundi\'c et al., determine
the delay with high precision, uncertainties in the lens model at present
limit the accuracy of a distance determination in this system
(\cite{falco91}; Grogin \& Narayan 1996a\&b;
\cite{bernstein97}; \cite{barkana99}; \cite{bernstein99}).
Recently time delays have been reported in other lens systems,
including PG~1115+080 (\cite{schechter97}), PKS~1830-211
(\cite{lovell98}), HE~1104-1805 (\cite{wisotzki98}), 
B1608+656 (\cite{fassnacht00}), and B1600+434 (\cite{koopmans00}).
For most of these lenses as well, the modeling currently limits the
determination of angular diameter distance.
In order to refine significantly gravitational lens investigations of cosmography,
it will probably be necessary to consider measurements of many lens
systems as well as to improve individual lens models.
With this goal in mind a number of lens monitoring programs are under way.

\begin{figure}
\vspace{10in}
\hspace{-1.25in}
\includegraphics{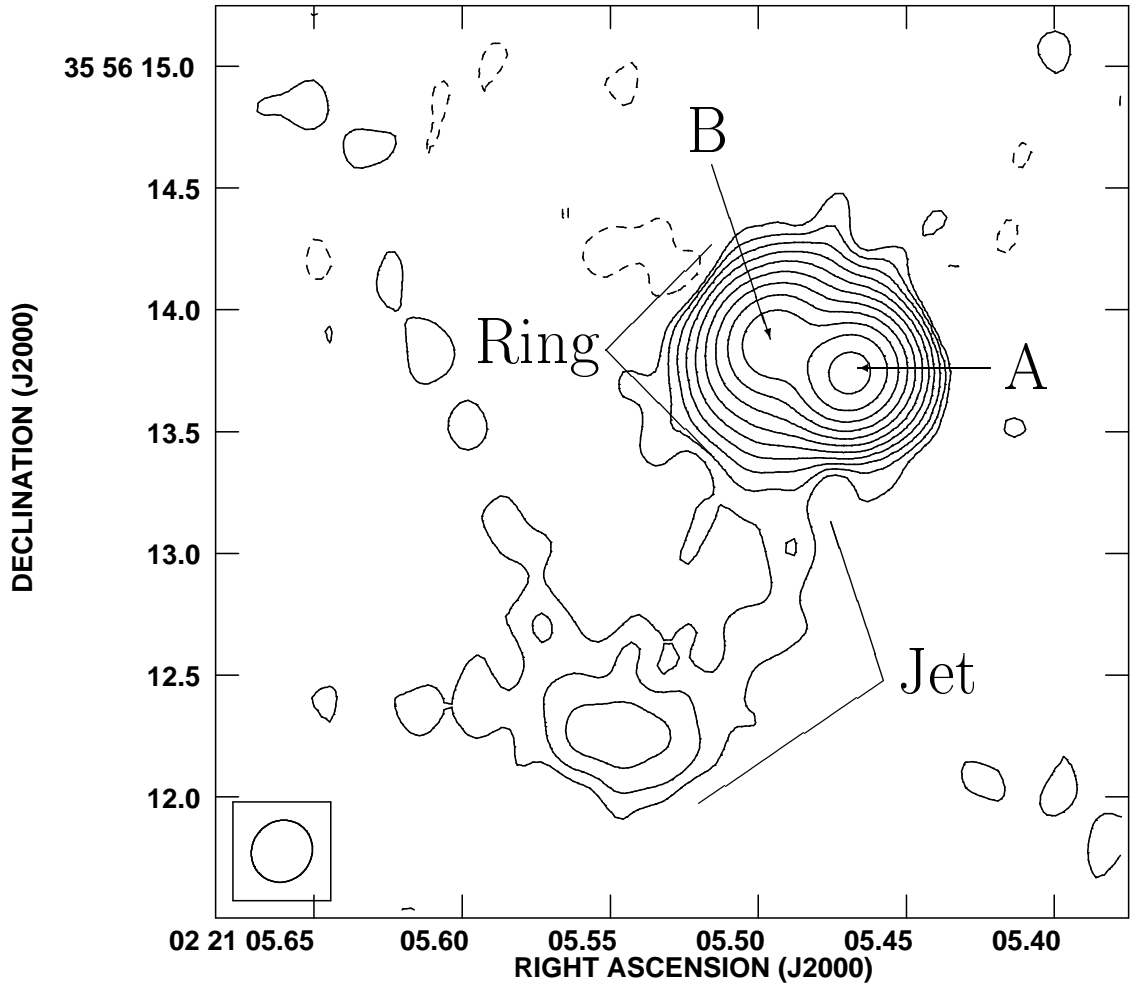}
\hspace{1in}
\vspace{-3in}
\caption{VLA 8 GHz map of 0218+357.  Contour levels are (0.6 mJy/beam)$\times$(-4, -2, -1, 1, 2, 4, 8, 16, 32, 64, 128, 256, 512, 1024, 2048). The peak brightness is
836 mJy/beam, and the beam major and minor axes are
263 and 250 milliarcseconds at a position angle of $3^\circ$.}
\label{xmap.fig}
\end{figure}

\begin{figure}
\vspace{10in}
\hspace{-1.25in}
\includegraphics{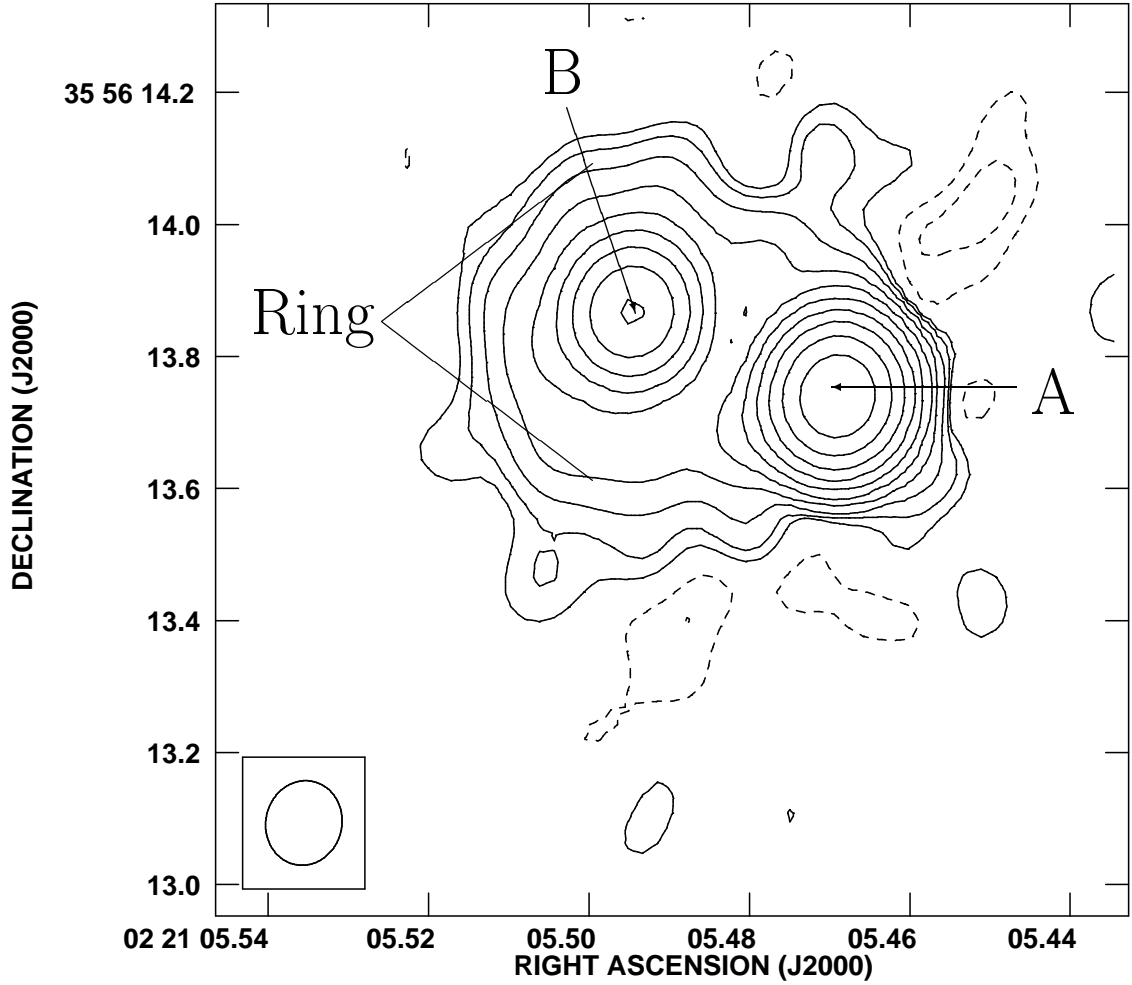}
\vspace{-3in}
\caption{VLA 15 GHz map of 0218+357.  
Contour levels are (0.8 mJy/beam)$\times$(-4, -2, -1, 1, 2, 4, 8, 16, 32, 64, 128, 256, 512, 1024, 2048). 
The peak brightness is 775 mJy/beam, and the beam major and minor
axes are 174 and 131 milliarcseconds at a position angle of
$15^\circ$.}
\label{umap.fig}
\end{figure}

The radio source 0218+357 was first proposed as a gravitational lens
by Patnaik et al. (1993) and O'Dea et al. (1992).
It consists of two images of a background BL~Lac object, including a
radio Einstein ring, 
and an extended jet and lobe emission  that are not multiply
imaged (see Figures \ref{xmap.fig} and \ref{umap.fig}).
The radio ring has an unusually small diameter, only $0.35''$.
The small ring, the detection of absorption lines attributed to neutral
hydrogen (\cite{carilli93}) and various molecules
(\cite{wiklind95}), the large rotation measure observed in 
linear polarization, and the lens optical colors and 
redshift  (\cite{keeton98})
suggest that the lensing object is a gas-rich
spiral galaxy.
The images of the radio core reveal a flat-spectrum component that has 
been imaged on VLBI scales (\cite{patnaik95}, \cite{patnaik99}).
The lens redshift, based on optical (\cite{browne93}) and radio
absorption lines, is 0.685; the likely source redshift is 0.96 
(\cite{lawrence96}).

VLA monitoring observations of 0218+357 have been carried out by 
Biggs et al. (1999), and they
report a time delay of $10.5\pm0.4$~days (95\% confidence).
They also report significantly different
flux ratios of $3.57\pm0.01$ and $3.73\pm0.01$ (68\% confidence)
at 8~GHz and 15~GHz, respectively.
These results are based on data spanning approximately 100~days.
The experience of 0957+561 has taught us that parameter fitting on a relatively
small dataset may fail to be robust, with different analysis techniques
yielding conflicting results.  Delays based on a single feature may be
unreliable due to the effects of
unmodeled processes, such as scintillation or gravitational
microlensing, which may bias the results.
Therefore, repeated measurements and different analysis techniques should
be brought to bear on the gravitational lens monitoring experiments.
The analysis of Biggs et al. included features found in the polarization
light curves as well as the flux density light curves,
which in effect represent repeated mesurements of the delay.
We report in this paper an investigation of
independent VLA measurements of the flux density of 0218+357, 
in which we used
data reduction and parameter fitting techniques that differed from those
of Biggs et al.
We also adopt a more general model of the source variability, including
the possibility of constant and variable components with different magnification
ratios.
A comparison of the results addresses the issue of the robustness of the
parameter estimates.
We caution, however, that since our monitoring observations occured at
the same epoch as those of Biggs et al., 
unmodeled scintillation or microlensing
may be biasing both mesurements.

\section{Observations and Data Analysis}

We present a set of
60 observations carried out with the National
Radio Astronomy Observatory (NRAO) Very Large Array (VLA)
telescope.\footnote{The National Radio Astronomy Observatory is operated
by Associated Universities, Inc., under cooperative agreement with the 
National Science Foundation.}
Continuum observations (with 100~MHz bandwidth) were carried out between
1996 October~9 and 1997 January~14 in the A~configuration at both
8~GHz and 15~GHz.
The flux density scale was set by observations of 3C48, and observations of
0205+322 were used to compute complex antenna gains.
Both senses of circular polarization were recorded. 
The time on source per observation
varied, but the averages were 5 and 10 minutes for
the 8~GHz and 15~GHz observations, respectively.
The theoretical thermal map noise for the average
observation time is 64~$\mu$Jy/beam
at 8~GHz, and 170~$\mu$Jy/beam at 15~GHz. 
However, as we discuss below, the error in measuring the flux densities
of the two points sources in the map is dominated by other effects.

The data were calibrated using the NRAO's
Astronomical Image Processing Software (AIPS) software.
At each frequency a reference model
based on a CLEAN deconvolution 
(\cite{hogbom74}, \cite{clark80})  was adopted.
For the reference models, data taken on 1996 November~9 (8~GHz)
and 1996 November~8 (15~GHz) were used.
Phase corrections with respect to this model were computed, and
the AIPS task IMAGR was used to create naturally weighted maps and to
deconvolve (again using the CLEAN algorithm) the maps.
The AIPS task CALIB was used to self-calibrate the phases with respect
to the CLEAN components of each map, restoring the flux densities to
the values appropriate to the data, minimizing the bias
of the reference model.
The process of CLEANing and self-calibration was carried out until 
the flux densities of the A and B compact components converged,
usually requiring three iterations.
The typical synthesized beam full width at half maximum was $0.24''$ at 8~GHz
and $0.14''$ at 15~GHz, barely resolving the $0.334''$ separation of
A and B at 8~GHz.
Each UV data set was rotated, shifted and scaled so that in the resulting
map each of the two point sources fell exactly at the center of an image
pixel.
We estimated the flux density of A and B as the value of the flux density
at these two image pixels.
Typical maps are displayed in Figures~\ref{xmap.fig} and \ref{umap.fig}.

In contrast to the method used by Biggs et al., we did not attempt
to fit the UV data to a model, because the precise structure of the
ring and jet is uncertain and may be too complicated for approximation 
by a simple model.
Error in our technique is introduced by the fact that the size and
shape of the synthesized beam varies slightly from map to map.
One approach to remove this source of error might be to convolve all
clean components with a standard clean beam before restoring them to
the residual map.  
However, the different UV samplings in fact measure different parts of
the unknown structure, so with this approach unknown levels of residual
systematic error are likely to remain.
As discussed below we chose instead to
use our measurements of the source structure and the UV coverage of each
observation to compute the error and correct the flux
density measurements.  This technique has been shown to be extremely
effective in tests with the non-varying source MG0414+0534, giving 
flux density uncertainties dominated by flux density
calibration errors and as small as 1\% (Moore and Hewitt, submitted).
In 0218+357 the presence of substantial extended emission complicates
the analysis, but the correction of bias due to UV sampling is nonetheless
critical.  In any case,
our flux density estimation technique differs substantially from
that of Biggs et al., and comparison of our results will provide a
check for systematic error.

One source of error in the flux density estimates of each point source
is confusion from the diffuse emission from the ring
and confusion from the other point source.
The flux of the ring is presumed constant, and the contamination
from the other point source is reduced by at least a factor of 100
by its separation and the properties of the synthesized beam.
The confusion from the diffuse emission changes as the UV-coverage
changes.
We correct for this  by creating a synthetic image
based on a static model of the source.
The static model consists of a set of CLEAN components, obtained
from one high quality
image, which are Fourier transformed onto the points in
the UV plane that each individual observation sampled.
This synthetic dataset for each observation was reduced using
exactly the same methods as for the real data, and any resulting variations
in the flux densities are due only to the variations in the UV coverage.
These variations represent multiplicative correction factors which we
applied to the light curves.
This procedure produced light curves in which the flux densities
represent that of each point component plus some level of outside
contamination which should be essentially constant.  A constant contribution
to the flux density,
which includes the outside contamination as well as the constant
part of the point source flux density,
is a parameter for which we fit when we perform our final analysis
in Section~4.

A second source of error is in the flux calibration procedure.
Because 3C48 is heavily resolved at both frequencies, we constructed
a model for the source by combining all our 3C48 data.
We determined
our flux density scale over a limited UV range with respect to this model,
allowing a wider range of UV spacings to be used in the flux
calibration and increasing its reliability.
We also applied antenna gain corrections as a function of elevation
(R.~Perley, private communation).
Despite our precautions,
there is evidently still an overall flux density scale error which
changes A and B by the same factor in each observation.
Treating the phase calibrator, 0205+322, as a flux calibrator only
increased the scatter in the light curves, indicating 0205+322
is variable.
We therefore adopt the flux density calibration based on 3C48,
accepting a certain amount of correlated error,
which we identify and treat as such in our analysis.
We believe that the flux density scale is the dominant source
of error in our light curves.

\section{Results: The Light Curves}

\begin{figure}
\epsfbox{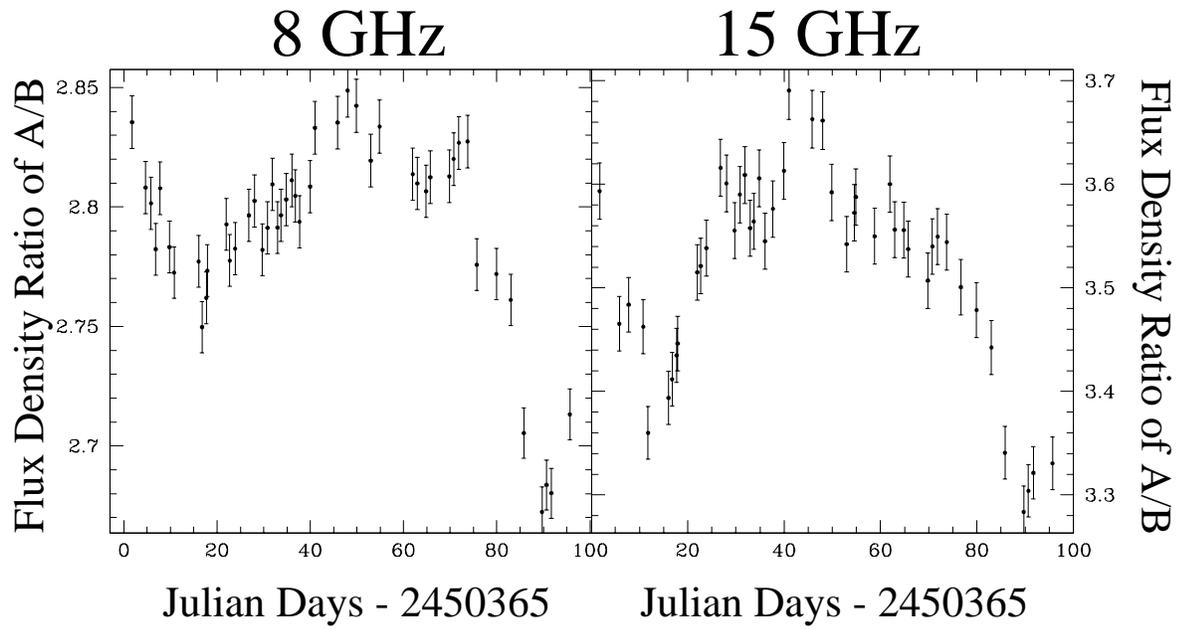}
\caption{Ratios of the A and B flux densities plotted
as a function of time.}
\label{ratiolc}
\end{figure}

In deriving our final light curves and their errors, we first
eliminated
some of the original 60 observations that were unreliable because of
bad weather.
We removed observations with reports of 
cumulus cloud cover of more than 50\%,
high wind, or precipitation.
That left 51 data points over a $\sim$100 day period for both
frequencies.
To estimate our measurement errors,
we used the smoothness of the light curves and took as our error estimate
the average
fractional difference between points on our light curve that are separated
by less than two days.
This assumes that within any two-day time period, the variations
are dominated by measurement errors and not actual source variability.
To the extent that this is not true, we can consider our result to
be an upper limit on the measurement errors.
For the 8~GHz light curves, the errors are 0.49\% and 0.62\% for 
A and B, respectively.
For the 15~GHz light curves, the errors are 1.4\% and 1.2\% for
A and B, respectively.
The relative levels of correlated and uncorrelated error were
determined by assuming they add in quadrature and by comparing
the A light curves with the flux ratios as a function of time
(see Figure~\ref{ratiolc}).
The error in the A light curve alone is the total measurement error,
while that in the A/B light curve represents only the uncorrelated
error, since the correlated errors divide out. 
We find 0.33\% and 0.99\% for the correlated errors at
8 and 15~GHz respectively.

\begin{figure}
\epsfbox{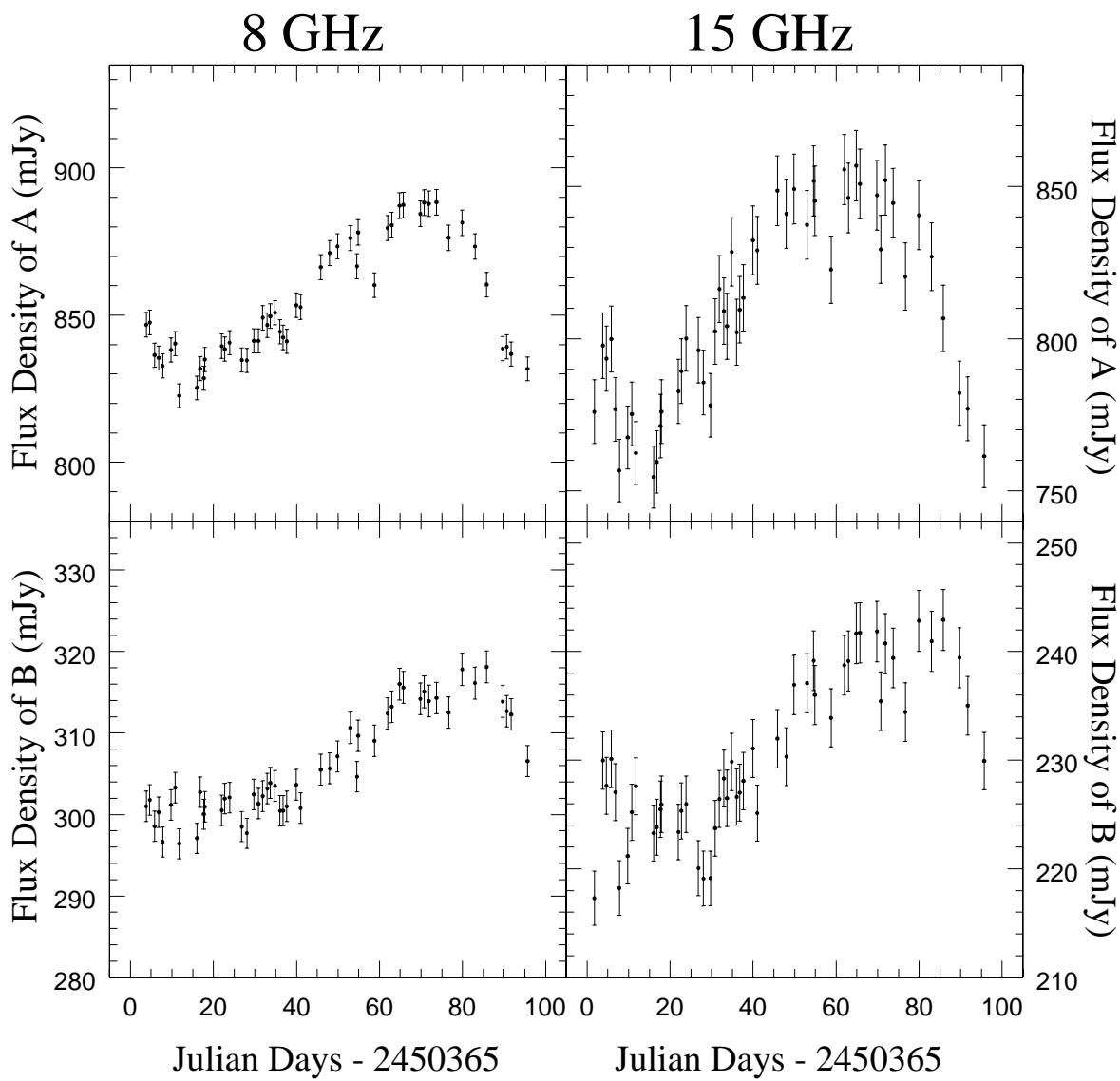}
\caption{Light curves of the
compact components of gravitational lens 0218+357.
These light curves include a constant offset due to a contribution
from the surrounding Einstein ring of radio emission.}
\label{lc.fig}
\end{figure}

Our final light curves are listed in Table \ref{lc.tab}
and plotted in Figure \ref{lc.fig}.
At both frequencies, there is a slow rise followed by a sharp
decline in the flux density of each component, in agreement with
the features seen by Biggs et al.
The sharp decline clearly occurs first in the A component, then
in the B component.
The fractional variability is greater in the 15~GHz light curves,
but the signal-to-noise ratio is greater in the 8~GHz light curves.
The A/B flux density ratios are apparently not the same at
both frequencies, in conflict with the achromatic nature
of gravitational lensing.
This discrepancy is probably primarily due to confusion
from the ring and the different resolution at the two frequencies.
It is also possible, and in fact likely,
that the A and B images are made up of
constant and varying components with different spectral
indices and magnification ratios
(see \cite{patnaik99}, \cite{press98}).
As discussed below, we take this into account by assuming a variability
model that consists of these two components, constant and varying.

\section{Results: Time Delays and Flux Ratios}

\subsection{Statistical Methods}

One can easily estimate the time delay between the light curves of
components A and B by eye, but it is preferable to have an objective
method that gives reproducible results and a quantitative estimate of the
errors.
Two methods are used in this paper.
The first is the maximum likelihood method of Press, Rybicki \& Hewitt (1992a,b;
henceforth PRH) and Rybicki \& Kleyna (1994), 
modified to account for the fact that some of flux
we are measuring could be constant and unrelated to the varying
part (Press \& Rybicki 1998).
The second method is the ``minimum dispersion method'' of Pelt et al.
(1994, 1996).

The first step in the PRH method is to determine the correlation properties
of the light curve as a function of time lag between measurements, where
we expect a greater correlation between measurements that have smaller
separations in time.
We fit a first-order structure function to
the data, modeling it as a power law:
$$
V(\tau) \equiv {1 \over 2} <[f(t) - f(t-\tau)]^2> = K \tau^\alpha
\eqno(1)
$$
where $f(t)$ a measurement at time $t$, $\tau$ is the time lag,
and $K$ and $\alpha$ are parameters that we fit 
(\cite{simonetti85}; \cite{edelson88}).
Since we do not know the flux ratio, we take $f(t)$ to be
the natural logarithm of the flux density measured at time $t$
referred to 1~mJy,
removing the dependence of our analysis on the unknown flux ratio.

As will be discussed in more detail in Section 6, we do not have enough
data to measure the instrinsic structure function ourselves with
a high degree of accuracy.
Therefore, we have to refer to other data and theory about extragalactic
variable sources in general to make reasonable assumptions.
In particular, long-term monitoring the flux density of many
quasars and BL~Lac objects (\cite{hughes92}) show that
for quasars, the structure function value of $\alpha$ is $1.04\pm0.18$
for light curves which are not dominated by a single feature (which tends to
give an upward bias on the measurement of $\alpha$).
For BL~Lac objects, $\alpha=0.94\pm0.37$.
There are also theoretical arguments that are consistent with this;
the value of $\alpha$ is exactly unity for such natural random processes
as shot noise and random walk.
It is reasonable, therefore, to assume that $\alpha=1$ in the
intrinsic structure function, and we do so in our analysis.

The next step in the PRH method is to test whether a single light curve,
constructed by combining the A and B light curves according to trial
parameters, has the same statistical properties as the individual light curves.
The combining of the A and B light curves is done nominally according
to four parameters.
The first is the time delay $T$, which results in the points from
the B light curve being shifted back in time by the amount $T$.
The second parameter is the magnification ratio; when combining the
light curves the B points must be multiplied by a magnification ratio, $R$.
Since we allow for the fact that each component contains a constant
component as well as a varying component, we specify that $R$ be the
flux ratio (A/B) of the variable components only. 
That leaves the constant parts of A and B ($C_A$ and $C_B$) as our third
and fourth parameters.  However, as shown by Press and Rybicki (1998),
we cannot solve for both $C_A$ and $C_B$.
The parameter for which we can solve is $C_o = RC_B - C_A$, which
essentially tells us the difference in the constant flux densities of
A and B.
The resulting three {\it independent} parameters are fit so as to best match
the statistical properties of the recombined light curve to that of
the individual light curves as described by the structure function.

As discussed by PRH and Rybicki and Kleyna (1994), the degree of ``matching''
is quantified by their so-called ``Q-statistic''
$$
\boldmath{Q = y^{T}B^{-1}y - \ln {{\rm det} (B^{-1})}}
\eqno(2)
$$
where $y$ is the vector of flux density values for the recombined light curve
and $\boldmath{B}$ is the total covariance matrix as defined by PRH.
The elements of the covariance matrix describe the statistical properties
of the light curve.  If we separate the measurement errors, $e(t)$ from
the true flux density, $s(t)$, then the measured flux density is
$$
f(t) = s(t) + e(t),
\eqno(3)
$$
which implies that
$$
B_{ij} = <s^2(t)> - V(\tau) + <e(t_i)e(t_j)>.
\eqno(4)
$$
The first term on the right-hand side is simply the average square of the
flux density values and is the same for all $B_{ij}$.  The second term is the
structure function that we have estimated empirically.
The last term is zero, except for two cases.  The first case is when
$i=j$ and $<e(t_i)e(t_j)> = <e^2(t)>$, the average square of
the measurement errors.
The other case occurs when $i \neq j$ but (because of the shift by the
time delay) the two flux density measurements come from the same observations
and so the errors are not uncorrelated.  In this way we account for the
correlated flux density scale errors.
Combining the light curves for trial values of $T$, $R$, and $C_o$, we
find the best-fit values of these parameters by minimizing the Q-statistic.

The Pelt et al. minimum dispersion method is similar to the PRH method in
that it also attempts to find recombination parameters that minimize a statistical
quantitiy associated with the combined light curve.
The quantity in this case is the dispersion, 
which is defined as the average square
of the difference in flux density values of nearby points in the light curve.
``Nearby'' is defined as points that are spaced apart in time by less than
the decorrelation length, $\delta$.
The contribution of any such pair of points is then weighted by the
factor $(1-\tau/\delta)$ where $\tau$ is the time between the two
points (\cite{pelt96}).
Since we give all the points in the light curves equal weight, our dispersion
statistic is:
$$
D = {\sum_{i,j} S_{i,j}(f(t_i)-f(t_j))^2(1-{\tau \over \delta}) \over
\sum_{i,j} S_{i,j}(1 - {\tau \over \delta})}
\eqno(5)
$$
where
$$
\tau = |t_i - t_j| ~~~{\rm and}~~~ S_{i,j} \left\{
                     \begin{array}{ll}
                        = 1 & {\rm if~} \tau < \delta \\
                        = 0 & {\rm if~} \tau \geq \delta \\
                    \end{array} \right.
\eqno(6)
$$
As in the PRH method, we fit $T$, $R$, and $C_o$ so as to minimize the 
dispersion in the combined light curve.
It is unclear, however, what value to use for $\delta$.
Therefore, we tried a range of values of $\delta$ and found best-fit values
of the parameters as a function of $\delta$.

\subsection{Results of the PRH Method}

Empirical esimates of the A and B structure functions for the two frequencies
were formed by binning the 1275 pairs from the 51 data points (see Figure 
\ref{sf.fig}).
The finite time series do not provide fair samples of the larger time
lags, so only the first section of the structure functions (from 0 to
about 50 days) were used in the fit to the power law model.
Since we assume $\alpha=1$, we fit only for $K$.
For the purpose of the time delay analysis, structure functions were fit
to both the A and B light curves, and at each frequency the results
were averaged ($<K> = \sqrt{K_A K_B}$), giving $<K>=1.2 \times 10^{-5}$
at 8~GHz and $<K> = 3.1 \times 10^{-5}$ at 15~GHz.
Finally the light curve recombination parameters were fit as described
above.
The results were $T=9.60$~days, $R=3.15$, and $C_o=113$~mJy for 8~GHz;
and $T=11.3$~days, $R=4.25$, and $C_o=176$~mJy for 15~GHz.

\subsection{Results of the Minimum Dispersion Method}

\begin{figure}
\hspace{0.7in}
\epsfbox{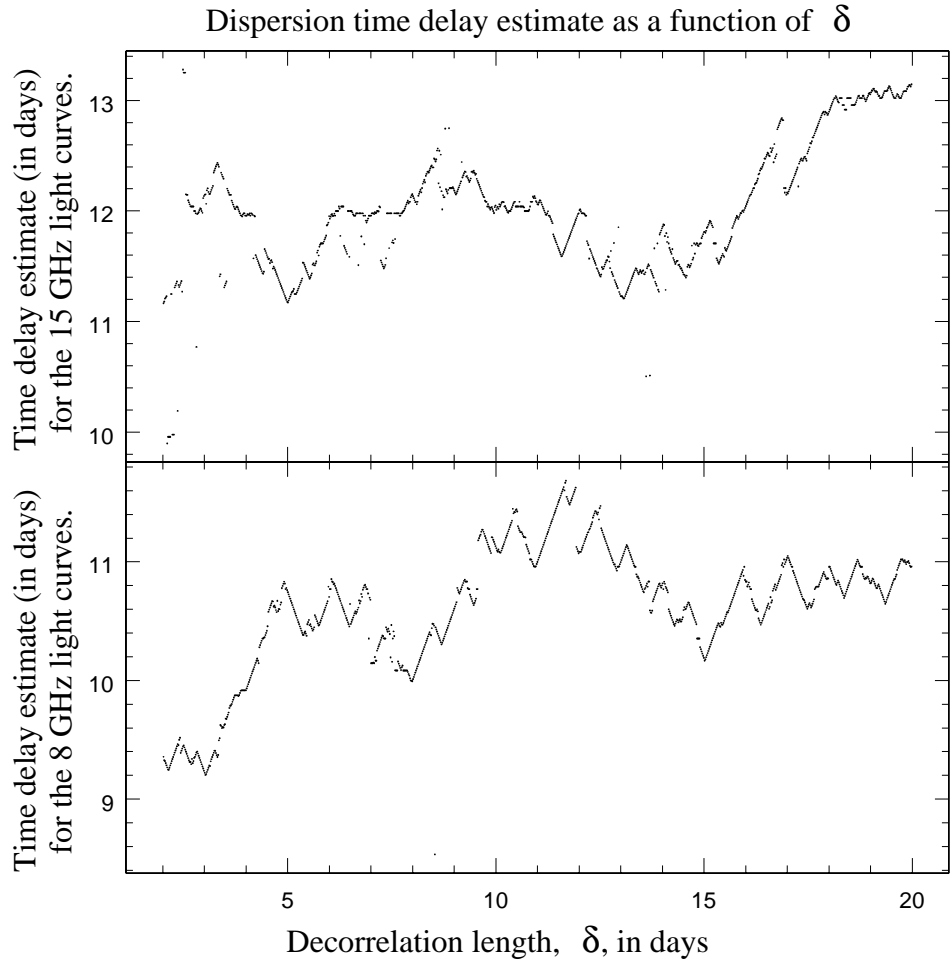}
\caption{The best fit time delay  and 
variable ratio as determined by the minimum dispersion method, plotted as a 
function of the ``decorrelation length'', $\delta$ (in days).  Since we 
have no apriori knowledge of $\delta$, this technique does not provide
definitive values for the parameters.}
\label{disp.fig}
\end{figure}

The results of the minimum dispersion method are shown in Figure 
\ref{disp.fig},
plotted as a function of the assumed decorrelation length $\delta$.
The results vary by $\pm 10$\% as a function of $\delta$,
even without any consideration of the errors in the parameters.
The values of $T$ are generally centered around the values
we found with the PRH method.
However, since we have no knowledge of the actual value of $\delta$ for
either light curve, we conclude that no definitive determination of the
parameters could be made with the dispersion method.
Although successful in determining the time delay in 0957+571 (Pelt et al.
1996), the method is inconclusive here, probably because in this case
we have a much shorter monitoring period.

\section{Error Analysis}

To estimate the accuracy  of our parameter fitting procedure, we
repeated the PRH analysis with two sets of simulated data.
The first set assumes that the underlying process producing the
light is gaussian; the second makes no such assumption and instead
derives the simulated data from the real data.

\subsection{Gaussian Monte Carlo Simulations}

As described by PRH, we created simulated light curves with the
same sampling in time as the true light curves, assuming a gaussian
process with the structure function and errors that we derived from
the measured light curves.
The measurement errors consisted of both  correlated and uncorrelated
parts, as described in Section~3.

\subsubsection{Errors Due to an Incorrect Structure Function}

Our first application of the Monte Carlo simulations was to investigate
the effect of an incorrect structure function on our estimate of
the time delay, $T$.
For each light curve, simulations were performed with the assigned time
delay varying randomly and uniformly between 0 and 20 days.
The differences between the actual time delays and the fitted time delays
were computed, and standard deviations of the set of differences were
calculated.
It is important to realize that for these calculations {\it two}
structure functions are involved.
The first is used to create the simulated data, and we call this
the ``true'' structure function.
The second is the one we fit to the simulated data in the process of
estimating the time delay; we call this the ``assumed'' structure
function.
Varying both the true and assumed structure functions in our simulations 
allows us to test the effects of an incorrect assumed structure function
on our time delay estimation, as a function of the true structure function.

Of the two free parameters in a structure function, $K$ and $\alpha$,
the value of $\alpha$ was chosen and the value of $K$ was fitted,
reproducing our analysis procedure.
For the ``true'' structure functions a ``true'' $\alpha$ was chosen and the
``true'' $K$ was fit to the real (measured with the VLA) light curves.
For the ``assumed'' structure functions an ``assumed'' $\alpha$ was chosen
and the ``assumed'' $K$ was fit to the simulated light curves.
Then the effectiveness of the PRH method was tested for values of
``true'' $\alpha$ and ``assumed'' $\alpha$ that ranged 
in five steps from 0.50 to 1.50.
For each case, we performed 1,000 Monte Carlo simulations.
The first result of this test showed that the average deviation between
the actual time delay and the PRH-deduced time delay was 
less than .02 days.
This was the case regardless of how the ``true'' and ``assumed'' 
values of $\alpha$ were varied. 
Therefore, it seems unlikely that an incorrect ``assumed'' structure function could 
have produced a signficiant bias in the time delay measurements.
However, the error estimates
 for each measurement did change somewhat as the structure
functions were varied.  Table \ref{alpha.tab} shows how the error 
estimates change
as function of ``true'' $\alpha$ and ``assumed'' $\alpha$.
The ``true'' $\alpha$ has much more of an effect on the error estimates  than the
``assumed'' $\alpha$.
This indicates that the magnitude of the error
estimates  depends mostly on the intrinsic
structure function of the BL~Lac object and very little on the accuracy of
the fitted structure function.  
Thus, the reliability of our error estimates appear to be limited
by our knowledge of the intrinsic structure function.
In all further simulations we assume $\alpha=1$.

\subsubsection{Confidence Intervals}

After settling on a reasonable structure function, we concentrate on our ability
to determine confidence intervals for the three parameters $T$, $R$, and $C_o$.
We expect $T$ and possibly
$R$ to be the same for both frequencies, but the $C_o$
parameter is likely to be different at the two frequencies.
We again constructed
Monte Carlo light curves, but now varying the assigned values of all
three parameters:
$T$ was varied between 8 and 12 days; $R$ was varied betweeen 2 and 5;
and $C_o$ was varied between 0 and 200~mJy.
For the light curves,
the fitted values were compared to the true values, and 95\% confidence
intervals were derived and adopted as the errors on the parameter estimates.
The distributions are shown in Figure \ref{hist.fig}.
The parameters $T$ and $R$ for the two frequencies were averaged,
weighted according to their errors.
The results are presented in Table \ref{mcresults}.

\subsection{Jackknife Samples}

\begin{figure}
\vspace{7.5 in}
\includegraphics{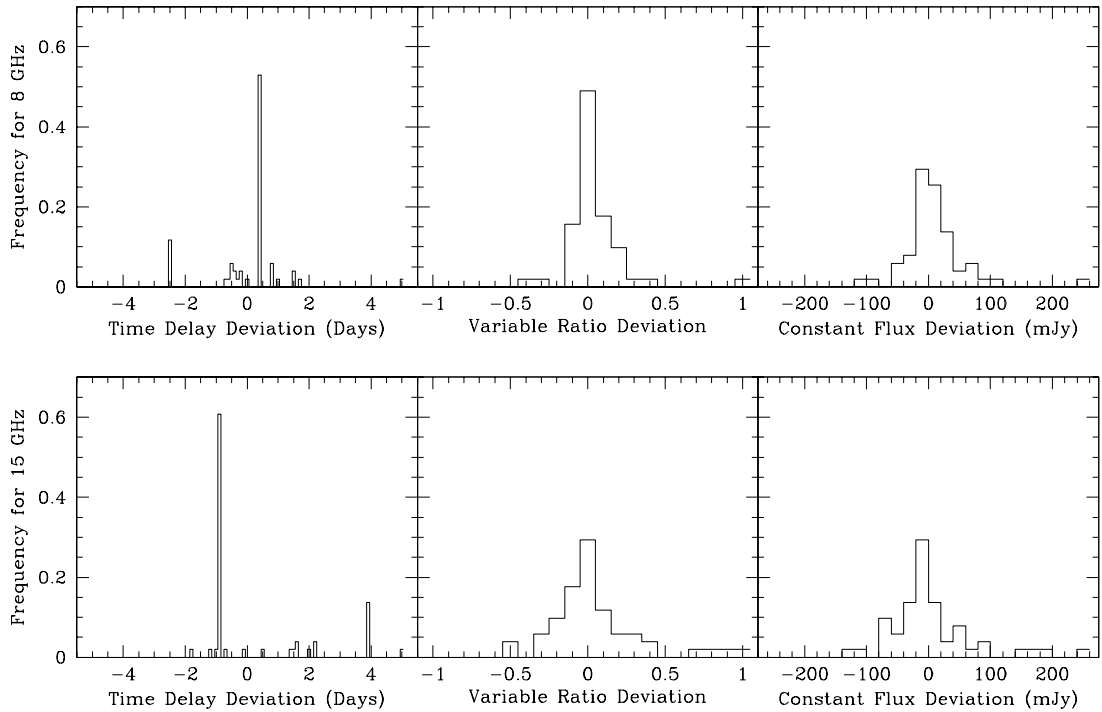}
\caption{Histograms that display the error distribution for each
parameter as deduced from jackknife samples of the 8~GHz and 15~GHz light
curves. 
These distributions include a rescaling of the horizontal axes by
the ``expansion factor''
$(N-1) / N^{1/2}$
(Efron \& Tibshirani 1993).}
\label{jkdists}
\end{figure}

The confidence intervals derived in the previous section are based on
the assumption that the underlying process producing the quasar light
curves is a gaussian process and on the model we assumed for the errors.
This is a weakness of our gaussian Monte Carlo technique, and we seek to 
explore methods that derive the statistics of the process from the
data themselves.
One such method is the jackknife (\cite{tukey58}; see also
\cite{efron93}), in which ``jackknife samples'' are formed by successively
deleting one point from the data set.
We applied this technique to our light curves at both frequencies,
estimating the $T$, $R$, and $C_o$ parameters for each jackknife sample
and forming distributions of the fitted values, shown in Figure \ref{jkdists}.

We compute errors on the parameters by forming the 95\% confidence intervals
from the data of Figure \ref{jkdists}, which have been rescaled by
the necessary
``expansion factor'' $(N-1)/\sqrt{N}$ (see Efron \& Tibshirani 1993).
Values from the different frequencies for $T$ and $R$ are combined
as described above, and the results are presented in Table \ref{jkresults}.
We caution that since the light curve data points are not independent,
the jackknife simulations are likely to underestimate the true errors.

\subsection{Evaluation of Errors}

\begin{figure}
\vspace{7.5in}
\hspace{-.25in}
\includegraphics{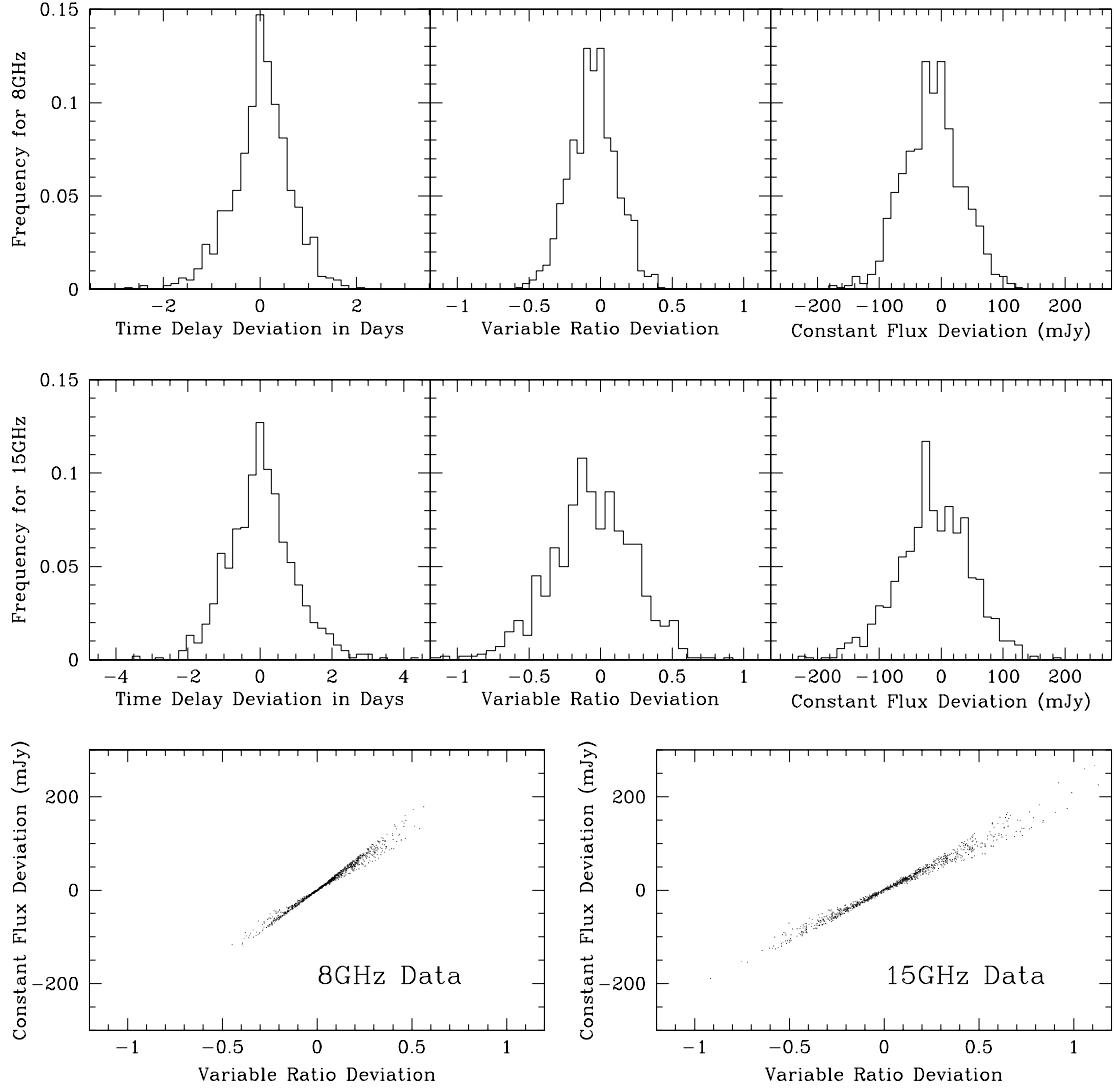}
\vspace{-2in}
\caption{Histograms that display the error distribution 
for each parameter as deduced from Monte Carlo simulations of
the 8 GHz and 15 GHz light curves.  
The ``deviation'' in each case is the difference between the fitted
and true values.
The bottom two panels compare the $C_o$ 
deviation to the  $R$ deviation for each simulated data set.  
There is clearly a high correlation between the two.  This demonstrates 
that if either $C_o$ or $R$ is known {\it a priori}, the other parameter is 
well constrained.  However, the two parameters  cannot be
constrained simultaneously with these data.}
\label{hist.fig}
\end{figure}

\begin{figure}
\epsfbox{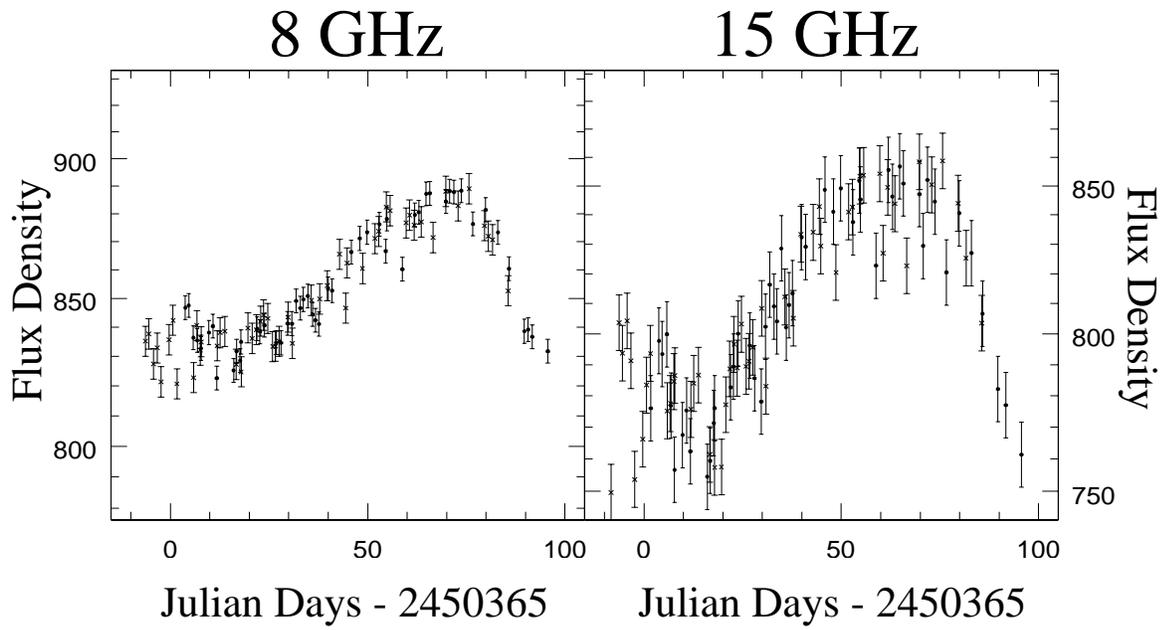}
\caption{Light curves of component A (filled
circles) and component B (X's)
superimposed according to the best-fit time delays
and magnification ratios.}
\label{comblc}
\end{figure}

If we compare the error distributions generated by the two different
simulation techniques (Figures~\ref{hist.fig} and \ref{jkdists}), 
a couple of interesting differences emerge.
First, the jackknife distribution for the time delays is very different
from the corresponding Monte Carlo distribution, and it clearly is not
gaussian.
This is an indication that the Monte Carlo simulations are failing to capture
the properties of the data in a way that causes estimation of the time
delay distribution to be unreliable.
Therefore, we adopt the jackknife errors as our errors on $T$.
Second, the jackknife distribution for $R$ and $C_o$ are not as broad as
the corresponding Monte Carlo distribution, but do appear gaussian,
indicating that
the $R$ and $C_o$ estimation process is better behaved than the $T$ 
estimation process.
However, since there is reason to suspect the jackknife procedure may
underestimate the true errors, we adopt the Monte Carlo errors as our
errors on $R$ and $C_o$.
The differences in the distributions in Figures~\ref{hist.fig} 
and \ref{jkdists}  illustrate the
difficulty of reliable error estimation in these light curves, a topic
that deserves further study.
With the confidence intervals determined, we now see that the
best-fit values of $T$ for the two observing frequencies
are not significantly different, and we form a weighted average.
The best-fit values of $R$ are only marginally significantly different, so
we also compute its average.
Our final results, with 95\% confidence intervals, are:
$T=10.1^{+1.5}_{-1.6}$~days, $R=3.2^{+0.3}_{-0.4}$ (8~GHz),
$R=4.3^{+0.5}_{-0.8}$ (15~GHz),
$R=3.4^{+0.2}_{-0.4}$ (averaged), 
$C_o = 110^{+80}_{-110}$ (8~GHz), and $C_o = 180^{+100}_{-140}$ (15~GHz).
We reiterate that the jackknife errors used for the time delay are
likely to be an underestimate.
The light curves superimposed according to the best-fit parameters are
shown in Figure~\ref{comblc}.

\section{Are the Structure Functions Significantly Different?}

\begin{figure}
\hspace{0.7in}
\vspace{5.5 in}
\includegraphics{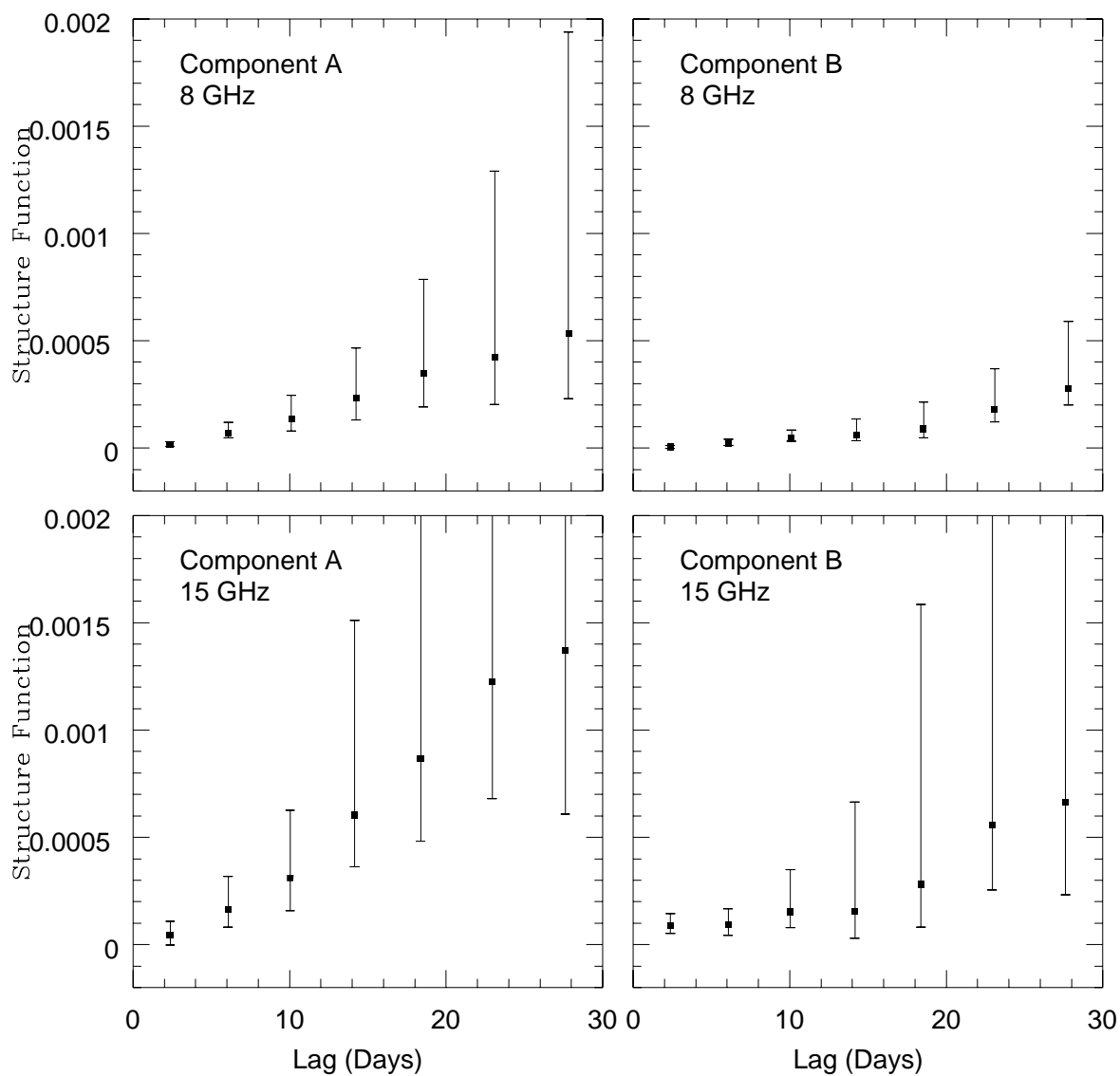}

\caption{Empirical point estimates of the structure function for 
the 8 GHz and 15 GHz light curves for A and B.  The points are binned 
in groups of 100.  
The error bars are derived from Monte Carlo simulations and show
that the differences in the structure functions are not significant.
Structure functions are computed according to equation (1), with flux densities
expressed in natural logarithm units referred to 1~mJy.}
\label{sf.fig}
\end{figure}

The A and B structure functions appear different (Figure~\ref{sf.fig}), 
with component B more
strongly auto-correlated at small lags than component A. Since radiation
from the two components travels along different paths to reach the
observer, such a comparison may reveal differences in propagation characteristics,
such as scintillation or gravitational microlensing.
In this section we investigate the significance of the apparent differences
in the structure functions.

The points plotted  in Figure \ref{sf.fig} represent bin averages.
If the (assumed white) light curve measurement noise is $\sigma_m$
and there are $N$ points in a bin, then the standard error of
the structure function esimate in that bin is approximately
$$
\sigma_V = {2 \sigma_m \over \sqrt{N}} \sqrt{V(\tau)}
\eqno(7)
$$
(Simonetti, Cordes \& Heeschen 1985, corrected for our choice of scaling
of the structure function), assuming a long time series.
However, our time series is truncated after only 100 days, so we cannot
necessarily assume the lags are fairly sampled.  Therefore, we have addressed this
empirically by further Monte Carlo simulation.

We again remove the dependence of $V(\tau)$ on the
assumed flux ratio by working in natural logarithmic units,
referenced to 1~mJy.  A power law
structure function was fit to the data, but now allowing the parameter $\alpha$ to
vary and fitting separately for components A and B at the two frequencies.
The four resulting structure function
models were used to generate four sets of gaussian
Monte Carlo simulations, of 1000 realizations each, and structure functions
were computed for all the simulated data sets.
From these the mean and the 68\% confidence interval were computed for
each value of $\tau$.
The results are the error bars shown in Figure \ref{sf.fig}.
Although our error estimates are to some extent model dependent,
the differences in the structure functions are not significant.
Much longer time series would be necessary to explore propagation effects
in this way.

\section{Comparison with Previous Measurements}

Our time delay value of $10.1^{+1.5}_{-1.6}$ days (95\% confidence)
is consistent with the value
found by Biggs et al. and lends confidence to the robustness of the
time delay measurements.  
Our results are derived from an independent
set of data, very different data reduction techniques, very
different parameter fitting techniques, and more
general models for both the measurement errors and 
the variability of the background object.  The light
curves produced by the two analyses are in excellent agreement, showing
the same major feature.  However, this is also a weakness of the comparison.
It is possible that this feature is affected by microlensing or scintillation,
and both analyses of the data would be biased in the same way if this is
the case (see \cite{haarsma99} for a discussion of such bias in the
0957+561 lens system).  While there is no evidence for microlensing or
scintillation in the light curves of 0218+357, 
only measurements at another epoch can reduce this uncertainty.

Our error analysis results in confidence intervals for the time delay
that are somewhat larger than those of Biggs et al.
The errors on our flux density measurements are smaller than those
of Biggs et al. at 8~GHz, larger at 15~GHz, but in both
cases not much different.
Therefore, it appears the source of the difference in confidence intervals
is in the time delay fitting procedures rather than in the light curve
errors.
Since we parameterize our variability model differently, with three
parameters associated with each light curve rather than two, it is not
surprising that our confidence intervals are larger.
In fact, we caution that in general
an oversimplification of the model may result
in deceptively small confidence intervals.

In the derivation of magnification ratios, Biggs et al. parameterized their
model in terms of the ratios of the total flux densities.  We
separate the variable and constant parts of the flux densities, fitting
for the magnification ratio associated with the variable part.
Biggs et al. find very different flux ratios at 8 and 15~GHz; the values
differ by more than ten standard deviations.
As discussed above, such differences may be caused by different source
structure at the different frequencies, combined with lens magnification
gradients and the nonzero resolution of the observations.
Our fitting procedure represents an attempt to derive the true magnification
ratio of the {\it variable part} of the source that presumably is small
in angular size.
The importance of this quantity is that the observational effects
that cause flux ratio differences at different frequencies should be
greatly reduced, and we are measuring the true lens magnification
that is important in constraining lens models.

The fitting of our more general model shows that for
the existing data there is a degeneracy between the $C_o$ and the
$R$ parameters, so that without independent information on $C_o$
the value of $R$ is rather poorly determined.  
Therefore, we find larger confidence intervals for the values of $R$
at the two frequencies, with the confidence intervals including the values
found by Biggs et al.
The experience of 0957+571 shows that a longer series of data that includes
both variable and quiescent behavior of the quasar should determine with
more precision the 
magnification ratio of the variable part of the source.

The implied value of the Hubble constant for
a particular lens model scales with the measurement of the time delay.
Therefore, applying our time delay value to the model of Biggs et al.
yields $H_{o} = 71^{+17}_{-23}\,{\rm km}\,{\rm s}^{-1}{\rm Mpc}^{-1}$ at 
$95\%$ confidence level, necessarily consistent with their result.
This model assumes a flat universe in which $\Omega_{o} = 1$ and 
$\lambda_{o} = 0$ (filled beam). 
Leh\'ar et al. (2000) have also attempted to model the 0218+357 lens, and
they conclude that the uncertainties are larger than those found
by Biggs et al.
We await the results of planned VLBA measurements of the radio ring
before tackling the problem of modeling the lens.
In any case, the above value of $H_o$ is consistent with other
recent measurements.
For example,  a value of $H_o = 71 \pm 6$~km~sec$^{-1}$~Mpc$^{-1}$
(67\% confidence interval) was determined by the HST key project
collaboration (\cite{mould99}), and a value of $H_o = 81\pm4$ 
(67\% confidence interval) was derived from a recalibration of the
Cepheid distance scale (\cite{maoz99}) based on a geometric distance to
NGC~4258 (\cite{herrnstein99}).
At this time, the time delay and mass model uncertainties are still
large enough to preclude any rigorous attempts to test
cosmological models with the 0218+357 lens system.

\section{Conclusions}

VLA flux density measurements for the lensed images of 0218+357
have been carried out with errors of 1\% or less, even
with only five to ten minutes of dwell time on source.
This precision is important in time delay measurements because
the variability of the background object is often limited.
We find a time delay of $10.1^{+1.5}_{-1.6}$ days (95\% confidence). 
We find magnification
ratios associated with the {\it variable} part of the light curves of
$3.2^{+0.3}_{-0.4}$ (8~GHz), $4.3^{+0.5}_{-0.8}$ (15~GHz), and
$3.4^{+0.2}_{-0.4}$ (averaged).  
We find best-fit values (which are rather
poorly determined) of the difference in the constant part of the light curves of
$110^{+80}_{-110}$~mJy (8~GHz) and $180^{+100}_{-140}$~mJy (15~GHz).
Our results, when compared to those of Biggs et al.,
indicate that the time delay measurement is very robust in this system.
The magnification is relatively poorly determined because it
is difficult to separate the variable and constant part of the light
curves; however, this quantity is the one properly associated with
the true magnification of the lens and is needed for lens modeling.
Our values for the magnifications at the two frequencies are different
at about the $3\sigma$ level.  However, given the uncertainties involved in
determining confidence intervals, we do not feel this is strong
enough evidence to claim frequency-dependent magnification ratios.
Monte Carlo simulations of structure function estimation have shown
that the structure functions are also poorly determined.
Thus, examining any differences in the variability properties of lensed
images, which would be interesting in the investigation of propagation
effects, would require much longer time series than are available in
this system at present.
We have demonstrated numerically that this uncertainty in the structure
functions does not significantly impact the time delay estimation,
as expected from theoretical considerations (Press, Rybicki and
Hewitt 1992a,b).
For the lens model of Biggs et al., the measured time delay implies
a Hubble constant of $H_o = 71^{+17}_{-23}$~km~s$^{-1}$~Mpc$^{-1}$
for $\Omega_o = 1$ and $\lambda_o=0$ (95\% confidence, filled beam). 
However, this confidence interval does not reflect systematic
error, which may be substantial, due to the uncertainty in the position
of the lens galaxy.

Gravitational lenses provide a way to determine the value 
of $H_{o}$ directly, without any dependence on intermediate distance
indicators.  This requires more data and modeling than currently exist.  
Of particular value would be to combine angular diameter
distance measurements for many
different lensed systems.  As more lenses are monitored for longer periods, 
the time delays inevitably will become far more accurate and numerous.  This 
leaves the lens models as the limiting factor in individual angular diameter
distance measurements.  Each lens is unique in this respect, but in general 
models can be improved by extensive observing at various frequencies and 
resolutions. 
Such measurements
improve the flux density and positional information about the 
multiply imaged source components and the lensing mass, and provide a 
greater number of modeling constraints.
In the lens system 0218+357, there is a full Einstein ring in addition to 
the doubly imaged radio core.  This ring is difficult to observe in detail
because it is small (about 350 mas in diameter) and very faint compared to 
the radio cores.  However, an accurate map of the fine structure in the 
ring will add many more constraints to any existing models of this system
and greatly improve their accuracy.  This is probably the 
best way to improve the measurement of the angular diameter distance to this 
lens.

\acknowledgments

We gratefully acknowledge the assistance of the VLA staff in the preparation
and execution of the extensive observations described in this paper.
This work was supported in part by grant AST-9617028.  A.S.C. 
acknowledges the support of an NSF Graduate Fellowship; D.B.H. 
acknowledges the support of a Cottrell College Science Award from
Research Corporation.
We thank the referee, A.~Biggs, for extensive comments that led to
improvements in the manuscript.

\begin{table}
\begin{scriptsize}
\begin{center}
\begin{tabular}{|c|c|c|c|c|c|}
\multicolumn{3}{c}{\large8 GHz Flux Densities} &
\multicolumn{3}{c}{\large15 GHz Flux Densities} \\
\tableline
Days since & Flux Density  & Flux Density & 
Days since & Flux Density  & Flux Density  \\ [-.05in]
Julian Day & of A & of B &
Julian Day & of A & of B \\ [-.05in]
2450365 & (mJy) & (mJy) &
2450365 & (mJy) & (mJy) \\
\tableline
3.81 & 846.7 & 301.0 & 1.75 & 776.0 & 217.3 \\ [-.1in]
4.67 & 847.5 & 301.8 & 3.81 & 797.7 & 230.0 \\ [-.1in]
5.83 & 836.4 & 298.6 & 4.67 & 793.4 & 227.6 \\ [-.1in]
6.83 & 835.5 & 300.3 & 5.83 & 799.9 & 230.1 \\ [-.1in]
7.77 & 832.8 & 296.6 & 6.83 & 776.8 & 227.0 \\ [-.1in]
9.79 & 838.2 & 301.2 & 7.77 & 756.7 & 218.2 \\ [-.1in]
10.79 & 840.4 & 303.3 & 9.79 & 767.5 & 221.2 \\ [-.1in]
11.77 & 822.6 & 296.4 & 10.79 & 775.3 & 225.2 \\ [-.1in]
16.04 & 825.3 & 297.1 & 11.77 & 762.4 & 227.6 \\ [-.1in]
16.79 & 831.8 & 302.7 & 16.04 & 754.5 & 223.3 \\ [-.1in]
17.69 & 828.6 & 300.0 & 16.79 & 759.5 & 223.8 \\ [-.1in]
17.92 & 835.0 & 301.0 & 17.69 & 771.3 & 225.5 \\ [-.1in]
22.00 & 839.5 & 300.5 & 17.92 & 776.0 & 225.9 \\ [-.1in]
22.71 & 838.5 & 302.0 & 22.00 & 782.7 & 223.4 \\ [-.1in]
23.88 & 840.7 & 302.1 & 22.71 & 789.3 & 225.3 \\ [-.1in]
26.83 & 834.7 & 298.5 & 23.88 & 800.1 & 226.0 \\ [-.1in]
28.04 & 834.6 & 297.7 & 26.83 & 796.2 & 220.1 \\ [-.1in]
29.73 & 841.3 & 302.4 & 28.04 & 785.6 & 219.1 \\ [-.1in]
30.83 & 841.2 & 301.3 & 29.73 & 778.1 & 219.1 \\ [-.1in]
31.88 & 849.1 & 302.3 & 30.83 & 802.4 & 223.7 \\ [-.1in]
32.96 & 846.6 & 303.2 & 31.88 & 816.3 & 226.4 \\ [-.1in]
33.71 & 849.6 & 303.9 & 32.96 & 809.1 & 228.3 \\ [-.1in]
34.83 & 850.8 & 303.5 & 33.71 & 804.1 & 226.5 \\ [-.1in]
36.04 & 844.4 & 300.4 & 34.83 & 828.5 & 229.9 \\ [-.1in]
36.77 & 842.4 & 300.5 & 36.04 & 802.1 & 226.6 \\ [-.1in]
37.67 & 841.1 & 301.0 & 36.77 & 809.5 & 227.0 \\ [-.1in]
39.96 & 853.4 & 303.6 & 37.67 & 813.4 & 228.1 \\ [-.1in]
41.00 & 852.7 & 300.8 & 39.96 & 832.3 & 231.1 \\ [-.1in]
45.85 & 866.3 & 305.5 & 41.00 & 829.1 & 225.1 \\ [-.1in]
48.00 & 871.1 & 305.7 & 45.85 & 848.7 & 232.0 \\ [-.1in]
49.88 & 873.4 & 307.1 & 48.00 & 841.1 & 230.3 \\ [-.1in]
52.98 & 876.2 & 310.6 & 49.88 & 849.2 & 236.9 \\ [-.1in]
54.58 & 866.6 & 304.6 & 52.98 & 837.5 & 237.1 \\ [-.1in]
54.88 & 878.1 & 309.7 & 54.58 & 851.9 & 239.2 \\ [-.1in]
58.79 & 860.2 & 309.1 & 54.88 & 845.3 & 236.0 \\ [-.1in]
61.96 & 879.6 & 312.4 & 58.79 & 822.7 & 233.9 \\ [-.1in]
62.96 & 880.6 & 313.2 & 61.96 & 855.6 & 238.7 \\ [-.1in]
64.83 & 887.2 & 316.0 & 62.96 & 846.3 & 239.1 \\ [-.1in]
65.75 & 887.4 & 315.6 & 64.83 & 856.9 & 241.7 \\ [-.1in]
69.83 & 884.4 & 314.2 & 65.75 & 850.9 & 241.7 \\ [-.1in]
70.75 & 888.2 & 315.1 & 69.83 & 847.2 & 241.8 \\ [-.1in]
71.85 & 887.8 & 313.9 & 70.75 & 829.4 & 235.4 \\ [-.1in]
73.73 & 888.4 & 314.3 & 71.85 & 852.1 & 240.7 \\ [-.1in]
76.67 & 876.3 & 312.5 & 73.73 & 844.6 & 239.4 \\ [-.1in]
79.92 & 881.4 & 317.8 & 76.67 & 820.4 & 234.4 \\ [-.1in]
83.00 & 873.3 & 316.1 & 79.92 & 840.6 & 242.8 \\ [-.1in]
85.81 & 860.4 & 318.1 & 83.00 & 827.0 & 240.9 \\ [-.1in]
89.71 & 838.6 & 313.9 & 85.81 & 806.7 & 242.9 \\ [-.1in]
90.67 & 839.2 & 312.7 & 89.71 & 782.1 & 239.4 \\ [-.1in]
91.71 & 836.7 & 312.3 & 91.71 & 777.0 & 235.0 \\ [-.1in]
95.67 & 831.8 & 306.6 & 95.67 & 761.4 & 229.9 \\ 
\tableline
Error & $0.49\%$ & $0.62\%$ & Error & $1.4\%$ & $1.2\%$ \\  
\tableline
\end{tabular}
\end{center}
\end{scriptsize}
\caption{Flux density measurements. 
Error bars are not given for each 
point individually, but rather for each data set as a whole in the form of a 
constant percentage error calculated from the scatter in each light curve, as 
described in Section~3.  The estimated percentage errors are 
listed in the last row of the table.
These flux density values include a constant offset due to a contribution
from the surrounding Einstein ring of radio emission.}
\label{lc.tab}
\end{table}

\begin{table}
\begin{center}
\hspace{0.2in}
\begin{tabular}{|c|c|c|c|c|c|}
\multicolumn{6}{c}{\large{Error Estimates as a Function of ``True'' 
$\alpha$ and ``Assumed'' $\alpha$ }} 
\\
\tableline
``Fitted'' $\alpha$
& $\alpha_{true}$ = 0.5
& $\alpha_{true}$ = 0.75
& $\alpha_{true}$ = 1.00
& $\alpha_{true}$ = 1.25
& $\alpha_{true}$ = 1.50  
\\
\tableline
&$1\sigma$ = 0.42&$1\sigma$ = 0.42&$1\sigma$ = 0.48&$1\sigma$ = 0.59&$1\sigma$ = 0.82\\
0.5
&$2\sigma$ = 1.72&$2\sigma$ = 1.35&$2\sigma$ = 1.40&$2\sigma$ = 1.46&$2\sigma$ = 1.73\\
\tableline
&$1\sigma$ = 0.48&$1\sigma$ = 0.41&$1\sigma$ = 0.44&$1\sigma$ = 0.56&$1\sigma$ = 0.70\\
0.75
&$2\sigma$ = 1.80&$2\sigma$ = 1.23&$2\sigma$ = 1.32&$2\sigma$ = 1.40&$2\sigma$ = 1.80\\
\tableline
&$1\sigma$ = 0.60&$1\sigma$ = 0.42&$1\sigma$ = 0.42&$1\sigma$ = 0.54&$1\sigma$ = 0.69\\
1.00
&$2\sigma$ = 2.04&$2\sigma$ = 1.16&$2\sigma$ = 1.11&$2\sigma$ = 1.23&$2\sigma$ = 1.59\\
\tableline
&$1\sigma$ = 0.67&$1\sigma$ = 0.49&$1\sigma$ = 0.48&$1\sigma$ = 0.49&$1\sigma$ = 0.68\\
1.25
&$2\sigma$ = 2.29&$2\sigma$ = 1.42&$2\sigma$ = 1.22&$2\sigma$ = 1.30&$2\sigma$ = 1.68\\
\tableline
&$1\sigma$ = 0.98&$1\sigma$ = 0.67&$1\sigma$ = 0.26&$1\sigma$ = 0.55&$1\sigma$ = 0.61\\
1.50
&$2\sigma$ = 2.82&$2\sigma$ = 1.79&$2\sigma$ = 1.18&$2\sigma$ = 1.49&$2\sigma$ = 1.49\\
\tableline
\end{tabular}
\caption{Results of Monte Carlo simulations for various
``true'' and ``assumed'' structure functions.  In each case, the synthetic 
light curves were created using a structure function with a given 
$\alpha$ (the ``true'' $\alpha$).  A structure function was then 
fit to the synthetic light curves with the condition that $\alpha$ be 
fixed at a given value (the ``assumed'' $\alpha$).  This fitted structure 
function was then used to get the time delay.  The time delay derived from 
this technique was compared to the actual time delay in each case.  For 
each pair of true $\alpha$ and assumed $\alpha$, 1,000 synthetic light 
curves were produced.  In each case the accuracy of the time delay fitting 
was expressed as $1\sigma$ and $2\sigma$ error bars, given in units of days.}
\label{alpha.tab}
\end{center}
\end{table}

\begin{table}
\begin{center}
\begin{tabular}{cccc}
\multicolumn{4}{c}{Results with Monte Carlo Error Estimates (95$\%$ confidence)} \\
\tableline
$\nu$ & $T$ (days) & $R$ & $C_o$ (mJy)\\
\tableline
8 GHz & $9.6^{+1.3}_{-1.2}$ & $3.2^{+0.3}_{-0.4}$ & $110^{+80}_{-110}$ \\
\tableline
15 GHz & $11.3^{+2.0}_{-1.8}$ & $4.3^{+0.5}_{-0.8}$ & $180^{+100}_{-140}$ \\
\tableline
averaged & $10.1^{+1.1}_{-1.0}$ & $3.4^{+0.2}_{-0.4}$ &  \\
\tableline
\end{tabular}
\end{center}
\caption{Results of fitting the time delay ($T$), variable 
ratio ($R$) and excess constant component ($C_o$) to minimize the PRH$Q$ 
statistic for the 8 GHz and 15 GHz light curves.  
The error bars (95\% confidence) are determined from Monte Carlo simulations.
The last line 
of the table shows the averaged result from the two light curves.}
\label{mcresults}
\end{table}

\begin{table}
\begin{center}
\begin{tabular}{cccc}
\multicolumn{4}{c}{Results with Jackknife Error Estimates (95$\%$ confidence)} \\
\tableline
$\nu$ & $T$ (days) & $R$ & $C_o$ (mJy)\\
\tableline
8 GHz & $9.6^{+1.7}_{-2.6}$ & $3.2^{+0.2}_{-0.3}$ & $110^{+70}_{-80}$ \\
\tableline
15 GHz & $11.3^{+4.3}_{-2.0}$ & $4.3^{+0.4}_{-0.6}$ & $180^{+90}_{-150}$ \\
\tableline
averaged & $10.1^{+1.5}_{-1.6}$ & $3.4^{+.2}_{-.2}$ &  \\
\tableline
\end{tabular}
\end{center}
\caption{Results of fitting the time delay ($T$), variable 
ratio ($R$) and excess constant component ($C_o$) to minimize the PRH$Q$ 
statistic for the 8 GHz and 15 GHz light curves.  
The error bars (95\% confidence) are determined from jackknife samples.
The last line 
of the table shows the averaged result from the two light curves.}
\label{jkresults}
\end{table}

\end{document}